\newcommand{\CLASSINPUTbaselinestretch}{1.8}
\documentclass[onecolumn, draft, 12pt]{IEEEtran}
\usepackage{amsmath,cases}
\usepackage{amsthm}
\usepackage{amssymb}
\usepackage{cite}
\usepackage{amsfonts}
\usepackage{enumerate}
\usepackage[final]{graphicx}

\newcommand{\vect}[1]{\ensuremath{\mathbf{#1}}}

\newcommand{\SaS}[4]{\ensuremath{S_{#1}\left(#2,\, #3,\, #4 \right)}}

\newcommand{\re}[1]{\ensuremath{#1^{R}}}
\newcommand{\im}[1]{\ensuremath{#1^{I}}}

\newcommand{\D}[1]{\ensuremath{\;{\rm d}#1}}

\makeatletter

\newcommand{\Rmnum}[1]{\expandafter\@slowromancap\romannumeral #1@}
\makeatother

\begin{document}

\title{Space-Time Coding over Fading Channels with Stable Noise}
\author{Junghoon Lee and Cihan Tepedelenlio\u{g}lu, \emph{Member, IEEE}
\thanks{The authors are with the School of Electrical, Computer, and Energy Engineering, Arizona
State University, Tempe, AZ 85287, USA. (Email:
\{junghoon.lee,cihan\}@asu.edu).} }
\maketitle

\begin{abstract}
This paper addresses the performance of space-time coding over fading channels with impulsive noise which is known to accurately capture network interference. We use the symmetric alpha stable noise distribution and adopt two models which assume dependent and independent noise components across receive antennas. We derive pairwise error probability (PEP) of orthogonal space-time block codes (STBC) with a benchmark genie-aided receiver (GAR), or the minimum distance receiver (MDR) which is optimal in the Gaussian case. For general space-time codes we propose a maximum-likelihood (ML) receiver, and its approximation at high signal-to-noise ratio (SNR). The resulting asymptotically optimal receiver (AOR) does not depend on noise parameters and is computationally simple. Monte-Carlo simulations are used to supplement our analytical results and compare the performance of the receivers. 
\end{abstract}


\section{Introduction}
\IEEEPARstart{T}{he} additive Gaussian noise model has long been used because it produces simple and tractable mathematical models which are useful for gaining insight into the underlying behavior of communication systems. As the physical reality of most practical channels demonstrate much more sophisticated effects such as bursts and impulses, which arise as a consequence of man-made activity such as automobile spark plugs \cite{Middleton1977}, microwave ovens \cite{Kanemoto1998}, and network interference \cite{Sousa1992, Ilow1998, Yang2003, Hughes2000, Haenggi2009, Win2009}, the Gaussian noise model may not be accurate. Such environments are also observed in urban and indoor channels as well as underwater acoustic channels \cite{Spaulding1977, Middleton1999}. Therefore, impulsive noise which captures these physical effects should be considered. In such wireless environments, the performance is degraded both by fading and impulsive noise. To combat fading, antenna arrays are often used, giving rise to multi-input multi-output (MIMO) systems. Space-time coding has been used as one of the powerful diversity techniques in MIMO systems.

A number of performance analyses of STBC have been reported in the literature where the noise is Gaussian (see e.g., \cite{Tarokh1999, Gharavi2004, Zhang2005}). Recently, some works in the area of STBC in the presence of impulsive noise have also been reported. Performance of space-time diversity/coding for power line channels with Middleton Class-A noise model was studied by simulations in \cite{Giovaneli2002}. In \cite{Tepedelenlioglu2007} the code design criteria and the PEP upper bound were derived over a fading channel with Middleton Class-A noise. Subsequent work in \cite{Xudong2007} provided a closed-form expression for symbol error rate (SER) of orthogonal STBC (OSTBC) when the noise follows a Gaussian mixture model.

Symmetric $\alpha$-Stable (S$\alpha$S) distributions are an important class of noise distributions which can successfully model a number of impulsive noise processes. Studies \cite{Sousa1992, Ilow1998, Yang2003, Hughes2000, Haenggi2009, Win2009} show that, in a multi-user network with power-law path loss, the multiple access interference results in a S$\alpha$S distribution, when the interfering nodes are scattered according to a spatial Poisson point process (PPP). In \cite{Anxin2004}, the performance evaluation of a MIMO system in S$\alpha$S noise was performed by simulation with no closed-form expression for the error probability. Subsequent works in \cite{Win2009} and \cite{Rajan2010} provided closed form expressions for the bit error rate (BER) of linear diversity combining schemes for S$\alpha$S noise environments in single-input multi-output (SIMO) environments. In \cite{Niranjayan2009, Niranjayan2010}, the optimal linear receivers for S$\alpha$S noise were studied in SIMO systems. To the best of our knowledge there is no analysis of MIMO systems over fading channels with S$\alpha$S noise. To close this gap in the literature, our goal is to design receivers for, and analyze the effect of S$\alpha$S noise on space-time coded systems. While the receivers derived herein apply to all space-time codes, the (PEP-based) performance analysis holds for OSTBCs.

Throughout this paper, we use $(\cdot)^H$ for Hermitian, $(\cdot)^T$ for transpose, $\rm{diag}$$(\mathbf{x})$ for a diagonal matrix with elements of $\mathbf{x}$ along the diagonal, $\lVert\cdot\rVert$ for the Frobenius norm for matrices and Euclidean norm for vectors, $\lambda_i(\cdot)$ for the $i^{th}$ largest eigenvalue of a matrix, $\Re \lbrace \cdot \rbrace$ to denote the real part, $\Im \lbrace \cdot \rbrace$ to denote the imaginary part. Also, we use $E_{A,B}(C)$ to denote the expected value of the random variable $C$ with respect to the distributions of the random variables $A,B$. Finally, we write $f(x) = O(g(x))$ as $x \rightarrow a$ to indicate that $\limsup_{x \rightarrow a}| f(x) / g(x) | < \infty$.

\section{System Model}
\label{sec:System_Model}
We consider a wireless communication system where the transmitter is equipped with $N_{t}$ antennas and the receiver with $N_{r}$ antennas. We consider the following standard MIMO flat-fading channel model:
\begin{equation}
\label{eqn:system_model}
\mathbf{Y} = \sqrt{\rho}\mathbf{H}\mathbf{S} + \mathbf{W}
\end{equation}
where $\mathbf{Y}$ is the $N_{r}$$\times$$T_{s}$ received signal matrix, and $T_{s}$ is the length of the transmitted
data block; $\mathbf{H}$ is an $N_{r}$$\times$$N_{t}$ matrix, with independent and identical distributed (i.i.d.) circularly symmetric complex Gaussian entries with mean zero and variance 1; the average transmitted power at each transmitting antenna is denoted by the scalar $\rho$; $\mathbf{S}$ is the $N_{t}$$\times$$T_{s}$ transmitted data block, which is transmitted from a codeword set $\mathcal{S}$ with equal probability; $\mathbf{W}$ is the $N_{r}$$\times$$T_{s}$ additive impulsive noise matrix, with elements that have a S$\alpha$S distribution, as explained next.

We first introduce real valued S$\alpha$S random variables, which will later be used to define its complex counterpart used in this paper. A real valued (not necessarily symmetric) $\alpha$-stable random variable, $w \sim \SaS{\alpha}{\sigma}{\beta}{\mu}$ has a characteristic function given by \cite{Nikias_Shao, Nolan}
\begin{equation}
\label{eqn:general_char_eqn}
\varphi(t)= \exp\left\lbrace j \mu t - |\sigma t|^{\alpha}(1-j \beta \, {\rm sign}(t) \, \omega(t,\alpha))\right \rbrace   ,
\end{equation}
where
\begin{align}
\omega(t,\alpha) =
\begin{cases}
 \tan \left(\frac{\pi \alpha}{2} \right) &  \alpha \neq 1 \\
 -\frac{2}{\pi}\log|t| & \alpha = 1
\end{cases}
\; \; ,
\end{align}
and
\begin{align}
{\rm sign}(t) =
\begin{cases}
 t = 1 & \text{if } t > 0 \\
 t = 0 & \text{if } t = 0 \\
 t = -1 & \text{if } t < 0
\end{cases}
\; \; \;,
\end{align}
$\alpha \in (0,2]$ is the characteristic exponent, $\beta \in [-1,1]$ is the skew, $\sigma \in (0,\infty)$ is the scale and $\mu \in (- \infty, \infty)$ is the shift parameter. When $\beta = 0$, $w$ has a symmetric distribution about $\mu$. When $\beta = 0$ and $\mu = 0$, $w$ is a S$\alpha$S random variable. When $\alpha = 2$ and $\beta = 0$, $w$ is Gaussian, which is the only S$\alpha$S random variable with finite variance. Since the Gaussian case is widely studied, we focus on $\alpha \in (0,2)$ throughout. When $\sigma = 1$ and $\mu = 0$, $w$ is said to be standardized \cite[pp. 20]{Taqqu}. Any S$\alpha$S random variable $w \sim \SaS{\alpha}{\sigma}{0}{0}$ can be written as compound Gaussian, i.e., of the form $w=\sqrt{A}G$, where $A$ and $G$ are independent, with $A \sim S_{\alpha/2}\left(\left[\cos(\pi\alpha/4)\right]^{2/\alpha},1,0\right)$ is positive skewed $\alpha$-stable random variable and $G \sim S_{2}\left(\sigma,0,0\right)$ is Gaussian random variable with mean zero and variance $2\sigma^{2}$ \cite[pp. 38]{Nikias_Shao}, \cite[pp. 20]{Taqqu}.

Although a closed-form expression for the PDF of S$\alpha$S random variables exists only for a few special cases (e.g. Gaussian ($\alpha = 2$) and Cauchy ($\alpha = 1$)), asymptotic expansions for $\alpha \in (0,2)$ are well known as $w \rightarrow \infty$:
\begin{equation}
\label{eqn:PDF_low_mag}
f_{\alpha}(w) = \alpha(1+\beta) C_{\alpha} w^{-\alpha-1} + O(w^{-2\alpha-1})
\end{equation}
where the constant $C_{\alpha} := \Gamma(\alpha) \sin(\pi \alpha /2) / \pi$ \cite{Nolan}. Additionally, if $w \sim S_{\alpha}\left(\sigma,\beta,0\right)$, the complementary cumulative distribution function (CCDF) of $w$ satisfies the asymptotic relation as $\lambda\rightarrow\infty$:
\begin{equation}
\label{eqn:CCDF}
P(w>\lambda)=C_{\alpha}\sigma^{\alpha}(1+\beta)\lambda^{-\alpha}+O\left(\lambda^{-{2\alpha}}\right) \text{.}
\end{equation}

In the following, we will briefly introduce two noise models (Model \Rmnum{1} and \Rmnum{2}) which assume dependent and independent noise components across antennas. In both Model \Rmnum{1} and \Rmnum{2}, the $T_s$ columns of $\mathbf{W}$, $\mathbf{w}_1,...,\mathbf{w}_{T_s}$, in \eqref{eqn:system_model} are independent.

\begin{description}
\item [$\bullet$] Under Model \Rmnum{1}, we assume $\vect{w}_k:=[w_{1,k},w_{2,k},...,w_{N_r,k}]^T$ is a complex isotropic S$\alpha$S random vector, defined as
    \begin{equation}
    \label{eqn:noise_Model_1}
    \vect{w}_k=\sqrt{A_k}(\re{\vect{G}_k}+j \im{\vect{G}_k})
    \end{equation}
    where the scalar random variable $A_k \sim \SaS{\alpha/2}{\left[ \cos(\pi \alpha /4)\right]^{2/\alpha}}{1}{0}$ is independent of $\re{\vect{G}_k}$ and $\im{\vect{G}_k}$ which are Gaussian random vectors with i.i.d. elements which have mean zero and variance $\sigma^2$. This is a good assumption when the receiving antennas are influenced by the same physical process creating the impulse, thereby making the $A_k$ of each branch the same. This might, for example, be an accurate model for a multi-antenna system where the antenna elements spaced closely. Mathematically, it is not difficult to see that in this case $w_{1,k},w_{2,k},...,w_{N_r,k}$ will be statistically dependent \cite[pp. 83]{Taqqu}.

\item [$\bullet$] Under Model \Rmnum{2}, the $j,k$ element of $\mathbf{W}$ is given by
    \begin{equation}
    [\mathbf{W}]_{j,k}=\sqrt{A_{j,k}}(\re{G}_{j,k}+j \im{G}_{j,k})
    \end{equation}
    where $A_{j,k},\re{G}_{j,k}$ and $\im{G}_{j,k}$ are distributed as in Model \Rmnum{1}, but are i.i.d., and $[\mathbf{W}]_{j,k}$ is the $(j,k)$ element of matrix $\mathbf{W}$.
\end{description}

In both Model \Rmnum{1} and \Rmnum{2}, $w_{j,k}$ has a unity scale parameter ($\sigma=1$), since any scale is subsumed in $\rho$ in \eqref{eqn:system_model}. It can be shown that only the moments of order $\alpha$ or less exist for any S$\alpha$S random variable \cite[pp. 22]{Nikias_Shao}, as a result of which the conventional definition of SNR holds only for the Gaussian case ($\alpha = 2$). However, with a slight abuse of terminology, we will refer to $\rho$ as the SNR, even when $\alpha <2$, since $\rho$ quantifies the relative scale of the signal versus the noise.

\section{Receiver Design and Performance}
\label{sec:model_1}
We assume throughout that the channel $\mathbf{H}$ is known at the receiver. Under Model \Rmnum{1}, we start with the GAR for which $A_k$ are assumed known at the receiver at each time $k=1,...,T_s$. The GAR is optimal when $A_k$ are known, so that its performance can serve as a benchmark for any practical receiver that does not have this knowledge.

\subsection{Genie-aided Receiver}
\label{subsec:GAR_case2}
The GAR maximizes the posterior probability and hence minimizes the probability of error, when $\mathbf{H}$ and $A_1,...,A_{T_s}$ are known. In the following, we are going to derive the decoding rule. To express in matrix form, we define $\mathbf{A}=\rm{diag}$$\left(\left[1/\sqrt{A_1},...,1/\sqrt{A_{T_s}}\right]\right)$. Right multiplying \eqref{eqn:system_model} by $\mathbf{A}$, we obtain:
\begin{equation}
\label{eqn:GAR_eq_2b}
\mathbf{YA} = \sqrt{\rho}\mathbf{HSA} + \mathbf{WA}
\end{equation}
so that the product $\mathbf{WA}$ has i.i.d. $\mathcal{CN}(0,1)$ entries. Since the elements of $\mathbf{WA}$ are now white Gaussian and the codewords are equally likely, the optimal decision rule is to minimize the Euclidean distance:
\begin{equation}
\label{eqn:GAR_eq_3b}
\mathbf{\hat{S}}=\underset{\mathbf{S}}{\operatorname{argmin}} \lVert \mathbf{YA}-\sqrt{\rho}\mathbf{HSA}\rVert^2 \text{.}
\end{equation}

To express the PEP that $\mathbf{S}$ is transmitted and $\mathbf{S'}$ is received for the GAR in \eqref{eqn:GAR_eq_3b}, we follow the derivation in the Gaussian noise case and obtain,
\begin{equation}
\label{eqn:GAR_eq_4b}
P\left(\mathbf{S} \rightarrow \mathbf{S'}\vert\mathbf{H,A}\right)
=Q\left(\sqrt{\frac{\rho\lVert\mathbf{H}\left(\mathbf{S}-\mathbf{S'}\right)\mathbf{A}\rVert^2}{2}}\right) \text{.}
\end{equation}
Using \eqref{eqn:GAR_eq_4b} and Craig's representation of the $Q$ function,
\begin{equation}
\label{eqn:GAR_eq_5b}
Q(x)=\frac{1}{\pi}\int_0^{\frac{\pi}{2}}\exp\left(-\frac{x^2}{2\sin^2\theta}\right)\D{\theta} \text{,}
\end{equation}
\eqref{eqn:GAR_eq_4b} can be expressed as follows:
\begin{equation}
\label{eqn:GAR_eq_6b}
P\left(\mathbf{S} \rightarrow \mathbf{S'}\vert\mathbf{H,A}\right)
=\frac{1}{\pi}\int_0^{\frac{\pi}{2}}\exp\left(\frac{\rho\lVert\mathbf{H}\left(\mathbf{S}-\mathbf{S'}\right)\mathbf{A}\rVert^2}{4\sin^2\theta}\right)\D{\theta} \text{.}
\end{equation}
Taking expectation with respect to $\mathbf{H}$ and $\mathbf{A}$, we get
\begin{equation}
\label{eqn:GAR_eq_8b}
E_\mathbf{H,A}P\left(\mathbf{S} \rightarrow \mathbf{S'}\vert\mathbf{A}\right)
=\frac{1}{\pi}\int_0^{\frac{\pi}{2}}E_\mathbf{A}\left[\prod_{i=1}^{N_t}\left(\frac{1}{1+\frac{\rho}{4\sin^2\theta}\lambda_i(\mathbf{B})}\right)^{N_r}\right]\D{\theta}
\end{equation}
where $\mathbf{B:=(S-S')AA}^H\mathbf{(S-S')}^H$. Using \eqref{eqn:GAR_eq_8b}, we can show that the code design criterion under S$\alpha$S noise remains the same as the Gaussian noise case as follows. To obtain the maximum diversity order, we need $\mathbf{B}$ to be a full rank matrix for any realization of $\mathbf{A}$ in \eqref{eqn:GAR_eq_8b}. Since $\mathbf{A}$ is diagonal with nonzero diagonal elements, it is a full rank matrix. Therefore, if the codeword different matrix $\mathbf{S-S'}$ is full rank, $\mathbf{B}$ is guaranteed to be a full rank matrix.

When $\mathbf{S-S'}$ is square and unitary which is satisfied by e.g., the Alamouti code \cite{Tarokh1999b}, the eigenvalues satisfy $\lambda_i(\mathbf{B})=1/A_i$. Substituting in \eqref{eqn:GAR_eq_8b}, using the statistical independence of $A_i$, and taking expectation, we show in Appendix \ref{app:GAR} that, as $\rho \rightarrow \infty$
\begin{equation}
\label{eqn:GAR_eq_PEP}
P\left(\mathbf{S} \rightarrow \mathbf{S'}\right) =
\left[\underbrace{\left(\frac{1}{2\sqrt{\pi}}\frac{\Gamma\left(\frac{\alpha N_t+1}{2}\right)}{\Gamma\left(\frac{\alpha N_t}{2}+1\right)}\right)^{-\frac{2}{\alpha N_t}} \left(\frac{\Gamma\left(1+\frac{\alpha}{2}\right)\Gamma\left(N_r-\frac{\alpha}{2}\right)}
{\Gamma\left(1-\frac{\alpha}{2}\right)\Gamma(N_r)}4^\frac{\alpha}{2}\right)^{-\frac{2}{\alpha}}}_{=:G_{\rm{GAR}}(N_t,N_r,\alpha)} \; \; \; \rho\right]
^{-\frac{\alpha N_t}{2}}+O\left(\rho^{-\frac{\alpha}{2}(N_t+1)}\right) \text{.}
\end{equation}
Using $(G_c \cdot \rho)^{-G_d}$ expression to present PEP, we can define the diversity order, $G_d$, and the coding gain, $G_c$, from the PEP. The coding gain is defined as the amount that bit energy or signal-to-noise power ratio can be reduced under the coding technique for a given bit error rate. In \eqref{eqn:GAR_eq_PEP}, the $G_c$ is $G_{\rm{GAR}}(N_t,N_r,\alpha)$ and the $G_d$ is $\alpha N_t/2$. The implications of \eqref{eqn:GAR_eq_PEP} are interesting, because it suggests that the diversity order depends on the number of transmit antennas, $N_t$, and the noise parameter, $\alpha$. However, the number of receive antennas, $N_r$, does not contribute to the diversity order. This is due to the fact that the noise is not i.i.d. across antennas in Model \Rmnum{1}.

In order to investigate the behavior of the coding gain as a function of $N_r$, by differentiating the natural logarithm of the coding gain with respect to $N_r$, we get
\begin{equation}
\label{eqn:C_GAR_3}
\frac{\partial}{\partial N_r}\log G_{\rm{GAR}}=-\frac{1}{\alpha}\left[\psi\left(N_r-\frac{\alpha }{2}\right)-\psi\left(N_r\right)\right]
\end{equation}
where $\psi(x) := \frac{d\log\Gamma(x)}{dx}$ is the digamma function as defined in \cite[pp. 258-259]{Abramowitz72}. In \eqref{eqn:C_GAR_3}, since $\psi(x)$ is a monotonically increasing function for $x>0$, the term inside the brackets is negative $\forall \alpha \in (0,2)$. Therefore, the coding gain is a monotonically increasing function of $N_r$. So, even though $N_r$ does not contribute to diversity, it does improve the coding gain. Regarding the analysis of $G_{\rm{GAR}}(N_t,N_r,\alpha)$ in \eqref{eqn:GAR_eq_PEP} with respect to $N_t$, it is shown in Appendix \ref{app:CG_GAR} that the coding gain is a monotonically decreasing and convex function of $N_t$.

For Model \Rmnum{2} the GAR can also be derived by using the Hadamard product with $\mathbf{A}$ which is a matrix with $(j,k)$ element $1/\sqrt{A_{j,k}}$. However, its performance is not tractable.

\subsection{Minimum Distance Receiver}
The MDR, which is optimal over Gaussian noise minimizes the Euclidean distance:
\begin{equation}
\label{eqn:MDR_eq_1b}
\mathbf{\hat{S}}=\underset{\mathbf{S}}{\operatorname{argmin}}\lVert\mathbf{Y}-\sqrt{\rho}\mathbf{HS}\rVert^2 \text{.}
\end{equation}
Note that unlike the GAR in \eqref{eqn:GAR_eq_3b}, the MDR does not depend on $\mathbf{A}$. We now derive the PEP for the MDR.
Define $\mathbf{E:=H(S-S')/\lVert H(S-S') \rVert}$, and let $e_{j,k}$ be the $(j,k)$ element of $\mathbf{E}$. The PEP and its upper bound for the MDR are given by:
\begin{eqnarray}
\label{eqn:MDR_eq_2b}
P\left(\mathbf{S} \rightarrow \mathbf{S'}\vert\mathbf{H,A}\right)
&=& Q\left(\sqrt{\frac{\rho\lVert\mathbf{H}\left(\mathbf{S}-\mathbf{S'}\right)\rVert^2}
{2\sum_{k=1}^{T_s}A_k\sum_{j=1}^{N_r}\lvert e_{j,k}\rvert^2}}\right) \\
\label{eqn:MDR_eq_3b}
&\leq& Q\left(\sqrt{\frac{\rho\lVert\mathbf{H}\left(\mathbf{S}-\mathbf{S'}\right)\rVert}
{2 A_{max}\sum_{k=1}^{T_s}\sum_{j=1}^{N_r}\lvert e_{j,k}\rvert^2}}\right) \\
\label{eqn:MDR_eq_4b}
&=& Q\left(\sqrt{\frac{\rho\lVert\mathbf{H}\left(\mathbf{S}-\mathbf{S'}\right)\rVert^2}{2 A_{max}}}\right) \\
\label{eqn:MDR_eq_5b}
&=& \frac{1}{\pi}\int_0^{\frac{\pi}{2}}\exp\left(\frac{\rho\lVert\mathbf{H}\left(\mathbf{S}-\mathbf{S'}\right)\rVert^2}
{4 \sin^2\theta A_{max}}\right)\D{\theta}
\end{eqnarray}
where $A_{max}:=\max_k A_k$ is the maximum value among $A_1,...,A_{T_s}$. In \eqref{eqn:MDR_eq_4b} we used the fact that $\lVert\mathbf{E}\rVert=1$, and in \eqref{eqn:MDR_eq_5b} we used \eqref{eqn:GAR_eq_5b}. Taking expectation with respect to $\mathbf{H}$ and $A_{max}$, the following upper bound on the average PEP is obtained:
\begin{equation}
\label{eqn:MDR_eq_7b}
E_{\mathbf{H},A_{max}}P\left(\mathbf{S} \rightarrow \mathbf{S'}\vert A_{max}\right)
\leq\frac{1}{\pi}\int_0^{\frac{\pi}{2}}E_{A_{max}}\left[\prod_{i=1}^{N_t}\left(\frac{1}{1+\frac{\rho}{4\sin^2\theta}\frac{\lambda_i(\mathbf{C})}{A_{max}}}\right)^{N_r}\right]\D{\theta}
\end{equation}
where $\mathbf{C:=(S-S')}\mathbf{(S-S')}^H$. When $\mathbf{S-S'}$ is square and unitary, we can rewrite \eqref{eqn:MDR_eq_7b} as follows:
\begin{equation}
\label{eqn:MDR_eq_8b}
E_{\mathbf{H},A_{max}}P\left(\mathbf{S} \rightarrow \mathbf{S'}\vert A_{max}\right)
\leq\frac{1}{\pi}\int_0^{\frac{\pi}{2}}E_{A_{max}}\left[\left(\frac{1}{1+\frac{\rho}{4\sin^2\theta}
\frac{1}{A_{max}}}\right)^{N_rN_t}\right]\D{\theta} \text{.}
\end{equation}
Taking expectation with respect to $A_{max}$, we show in Appendix \ref{app:MDR} that, as $\rho \rightarrow \infty$
\begin{equation}
\label{eqn:MDR_eq_PEP}
P\left(\mathbf{S} \rightarrow \mathbf{S'}\right) \leq
\left[\underbrace{\left(\frac{N_t}{2\sqrt{\pi}}\frac{\Gamma\left(\frac{1+\alpha}{2}\right)\Gamma\left(N_rN_t-\frac{\alpha}{2}\right)}
{\Gamma\left(1-\frac{\alpha}{2}\right)\Gamma(N_rN_t)}4^{\frac{\alpha}{2}}\right)^{-\frac{2}{\alpha}}}_{=:G_{\rm{MDR}}(N_t,N_r,\alpha)} \; \; \; \rho\right]^{-\frac{\alpha}{2}}+O\left(\rho^{-\alpha}\right) \text{.}
\end{equation}
Equation \eqref{eqn:MDR_eq_PEP} suggests that the diversity order is always $\alpha/2$ regardless the number of antennas which is reduced compared to the GAR where it was $\alpha N_t/2$.

The behavior of the coding gain as a function of $N_r$ can be obtained from the derivative given by
\begin{equation}
\label{eqn:C_MDR_1}
\frac{\partial}{\partial N_r}\log G_{\rm{MDR}}=-\frac{2N_t}{\alpha}\left[\psi\left(N_rN_t-\frac{\alpha }{2}\right)-\psi\left(N_rN_t\right)\right] \text{.}
\end{equation}
In \eqref{eqn:C_MDR_1}, since $\psi(x)$ is a monotonically increasing function for $x>0$, we can verify the term inside the brackets is negative $\forall \alpha \in (0,2)$. Therefore, the coding gain is a monotonically increasing function of $N_r$. Next, by differentiating the log-coding gain with respect to $N_t$, we get
\begin{equation}
\label{eqn:C_MDR_3}
\frac{\partial}{\partial N_t}\log G_{\rm{MDR}}=-\frac{2}{\alpha}\left[\frac{1}{N_t}+N_r\left[\psi\left(N_rN_t-\frac{\alpha}{2}\right)-\psi\left(N_rN_t\right)\right]\right] \text{.}
\end{equation}
It can be shown numerically that the $G_{\rm{MDR}}(N_t,N_r,\alpha)$ monotonically decreases with $N_t$ when $\alpha \in (0,\alpha_0)$ for some constant $\alpha_0$. Unlike the GAR, in case of the MDR the number of transmit antennas, $N_t$, does not contribute to the diversity order. Hence when $\alpha \in (0,\alpha_0)$ the performance of MDR will be worse as $N_t$ increases. Intuitively, the reason for the deterioration in performance is that when $\alpha$ is small, the sum of independent noise samples do not ``average out" like it does when the noise has a finite variance. In other words, when $\alpha$ is small enough the performance bound of MDR suffers from increased transmit antennas! On the other hand, the coding gain is a monotonically increasing function of $N_t$ when $\alpha \in (\alpha_1,2)$ for some constant $\alpha_1$. In other words, when $\alpha \in (\alpha_1,2)$ the coding gain increases as the number of transmit antennas increase. When $\alpha \in (\alpha_0,\alpha_1)$, the coding gain is a concave function of $N_t$. The values of $\alpha_0$ and $\alpha_1$ depend on $N_r$ (e.g. when $N_r=1$, $\alpha_0 \approx 1.333$ and $\alpha_1 \approx 1.799$).

For Model \Rmnum{2} the PEP of MDR can be derived by using $A_{max}:=\max_{j,k} A_{j,k}$ in \eqref{eqn:MDR_eq_3b}. Following the same derivation, the PEP of MDR for Model \Rmnum{2} is obtained by multiplying $G_{\rm{MDR}}(N_t,N_r,\alpha)$ in \eqref{eqn:MDR_eq_PEP} with $N_r^{-2/\alpha}$ which implies less coding gain and the same diversity order. This is in contrast with the GAR which will be shown in the simulations to have better performance under Model \Rmnum{2} compared to Model \Rmnum{1}. In conclusion, for S$\alpha$S noise environments the conventional MDR receiver has poor performance especially for small $\alpha$.

\subsection{Maximum Likelihood Receiver}
We introduce the optimal ML receiver for Model \Rmnum{1} and \Rmnum{2}. Firstly, the optimal ML receiver for Model \Rmnum{1} is given by
\begin{eqnarray}
\label{eqn:MLR_eq_2b}
\mathbf{\hat{S}}
&=&\underset{\mathbf{S}}{\operatorname{argmax}}\prod_{k=1}^{T_s}f_{\alpha}
\left(\lVert\mathbf{y}_k-\sqrt{\rho}\mathbf{Hs}_k\rVert\right) \\
\label{eqn:MLR_eq_3b}
&=&\underset{\mathbf{S}}{\operatorname{argmax}}\sum_{k=1}^{T_s}\log f_{\alpha}
\left(\lVert\mathbf{y}_k-\sqrt{\rho}\mathbf{Hs}_k\rVert\right)
\end{eqnarray}
where $f_{\alpha}(\lVert\mathbf{x}\rVert)$ is a probability density function of amplitude distribution of $d$-dimensional multivariate isotropic stable random variables and is given by \cite{Nolan2}:
\begin{equation}
\label{eqn:MLR_eq_4b}
f_{\alpha}(r)=\frac{2}{2^{d/2}\Gamma(d/2)}\int_0^{\infty}(rt)^{d/2}J_{d/2-1}(rt)e^{\sigma^{\alpha}t^{\alpha}}dt
\end{equation}
where $r=\lVert\mathbf{x}\rVert=\sqrt{X_1^2+ \cdots +X_d^2}$ and $J_\nu(\cdot)$ is the Bessel function of order $\nu$.

In case of Model \Rmnum{2}, complex symmetric $\alpha$-stable random variables are independent in both space and time. Thus, we can modify the optimal ML receiver for Model \Rmnum{2} as follows:
\begin{equation}
\label{eqn:MLR_eq_1a}
\mathbf{\hat{S}}=\underset{\mathbf{S}}{\operatorname{argmax}}\sum_{k=1}^{T_s}\sum_{j=1}^{N_r}\log f_{\alpha}
(\lVert[\mathbf{Y}]_{j,k}-\sqrt{\rho}[\mathbf{HS}]_{j,k}\rVert) \text{.}
\end{equation}
Since $f_{\alpha}(r)$ cannot be expressed in terms of closed-form elementary functions, these ML receivers are seen to be computationally complex, and dependent on the noise parameters $\sigma$ and $\alpha$. We now consider receivers that perform nearly optimally, with the advantage of reduced complexity and not requiring knowledge of noise parameters, when compared to the ML receivers.

\subsection{Asymptotically Optimal Receiver}
\label{subsec:AOR_case2}
To simplify \eqref{eqn:MLR_eq_2b}, we use the expression for the tail of $f_{\alpha}(\cdot)$ in \cite{Nolan2}
\begin{equation}
\label{eqn:AOR_eq_1b}
f_{\alpha}(r)= \alpha2^{\alpha}\frac{\sin(\pi\alpha/2)}{\pi\alpha/2}\frac{\Gamma((\alpha+2)/2)\Gamma((\alpha+d)/2)}{\Gamma(d/2)}
r^{-(\alpha+1)}+O(r^{-(2\alpha+1)})
\end{equation}
as $r \rightarrow \infty$, where we note that $\alpha2^{\alpha}\frac{\sin(\pi\alpha/2)}{\pi\alpha/2}\frac{\Gamma((\alpha+2)/2)\Gamma((\alpha+d)/2)}{\Gamma(d/2)} > 0$. Now, using the dominant term of \eqref{eqn:AOR_eq_1b} in \eqref{eqn:MLR_eq_2b} and simplifying, we get
\begin{eqnarray}
\label{eqn:AOR_eq_2b}
\mathbf{\hat{S}}
&=&\underset{\mathbf{S}}{\operatorname{argmin}}\prod_{k=1}^{T_s}\lVert\mathbf{y}_k-\sqrt{\rho}\mathbf{Hs}_k\rVert \\
\label{eqn:AOR_eq_3b}
&=&\underset{\mathbf{S}}{\operatorname{argmin}}\sum_{k=1}^{T_s}\log\lVert\mathbf{y}_k-\sqrt{\rho}\mathbf{Hs}_k\rVert \text{.}
\end{eqnarray}
Using same approach as Model \Rmnum{1}, we can modify the asymptotically optimal receiver for Model \Rmnum{2} as follows:
\begin{equation}
\label{eqn:AOR_eq_1a}
\mathbf{\hat{S}}=\underset{\mathbf{S}}{\operatorname{argmin}}\sum_{k=1}^{T_s}\sum_{j=1}^{N_r}
\log\lVert[\mathbf{Y}]_{j,k}-\sqrt{\rho}[\mathbf{HS}]_{j,k}\rVert \text{.}
\end{equation}
The resulting receivers are asymptotically optimal at high SNR and relatively simple.

A few comments about complexity of the ML receiver and AOR follow. In \eqref{eqn:MLR_eq_3b} and \eqref{eqn:AOR_eq_3b}, we need to evaluate matrix norms. The only difference between \eqref{eqn:MLR_eq_3b} and \eqref{eqn:AOR_eq_3b} is that the equation \eqref{eqn:MLR_eq_3b} needs to evaluate the metric in \eqref{eqn:MLR_eq_4b} additionally. In \eqref{eqn:MLR_eq_4b}, it is needed to evaluate an elementary function, a special function (i.e., the Bessel function) and an integration of these functions for each candidate codeword $\mathbf{S}$. Instead of evaluation of \eqref{eqn:MLR_eq_4b}, we can alternatively use a lookup table for the numerical values of \eqref{eqn:MLR_eq_4b}. Such a lookup table would have sizable memory requirements since a lookup table would contain values for each of the $\alpha$ and $\sigma$ values corresponding to the noise parameters. For example, if the sizes of quantized $\alpha$ and $\sigma$ values are $N_{\alpha}$ and $N_{\sigma}$ respectively, we need the $N_{\alpha} \cdot N_{\sigma}$ entries in the table. In addition to these kinds of high computational complexity, the ML receiver also requires to estimate $\alpha$ and $\sigma$ values of S$\alpha$S noise. However in case of the AOR which performs within a tenth of a dB of the ML receiver which will be shown in Section \ref{sec:Simulations}, we do not need to evaluate the equation \eqref{eqn:MLR_eq_4b} and estimate the $\alpha$ and $\sigma$ values.

Therefore we propose to use the AOR for impulsive noise due to its relatively low complexity and its reasonable performance. Though our analysis is based on the receivers for Model \Rmnum{1}, it will be similar in case of Model \Rmnum{2}. We note that the asymptotically optimal receivers in \eqref{eqn:AOR_eq_3b} and \eqref{eqn:AOR_eq_1a} are additive and therefore can be used in conjunction with the Viterbi algorithm when $\mathbf{S}$ is a codeword on a trellis.

\section{Simulations}
\label{sec:Simulations}
In this section, we verify our results through Monte Carlo simulations. In our simulations, we assume that $\alpha=1.43$, which corresponds to the value estimated in \cite{Nassar2008} for modeling radio frequency interference in laptop receivers. We also consider a ``highly impulsive" scenario, with $\alpha=0.5$, which corresponds to a path loss exponent of $2/\alpha=4$ in an environment where the interfering nodes are scattered according to a PPP on a two-dimensional plane \cite{Win2009}.

\subsection{Performance Results under Model \Rmnum{1}}
\begin{figure}[tb]
\begin{minipage}{1\textwidth}
\centering
\begin{center}
\includegraphics[height=8.5cm,keepaspectratio]{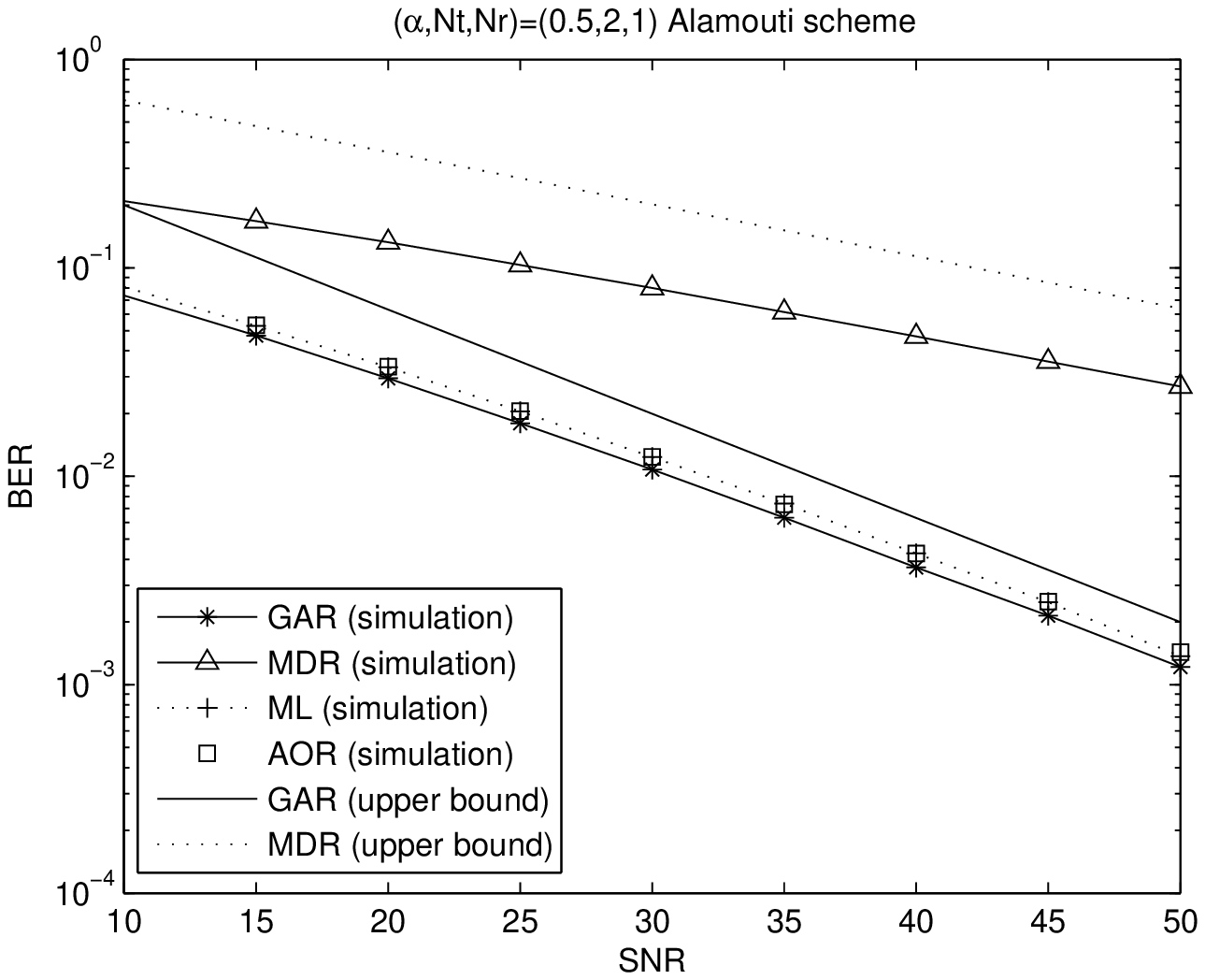}
\caption{Performance comparison of GAR, MDR, ML receiver, and AOR over a channel with highly impulsive noise ($\alpha = 0.5$) with $N_t=2$ and $N_r=1$}\label{fig:Alamouti_2x1_H}
\end{center}
\end{minipage}
\end{figure}

We show in Fig. \ref{fig:Alamouti_2x1_H} the performance bound of GAR for Alamouti code with $N_t=2,N_r=1$ over highly impulsive noise with BPSK. We calculate the BER union bound using the PEP of GAR in \eqref{eqn:GAR_eq_PEP}. We also plot the upper bounds for the MDR obtained using \eqref{eqn:MDR_eq_PEP}. In Fig. \ref{fig:Alamouti_2x1_H}, we also show the simulated BER results of Alamouti code for GAR, MDR, ML receiver, and AOR. Comparing  between theoretical and simulated results, we observe the diversity orders of GAR and MDR are $\alpha N_t/2$ and $\alpha/2$. We also observe the performance gap between ML receiver and GAR is about 1.3 dB at $10^{-2}$ BER. We also found the performance for AOR which does not need the noise parameters shows a difference less than a tenth of a dB to the ML receiver.

\begin{figure}[tb]
\begin{minipage}{1\textwidth}
\centering
\begin{center}
\includegraphics[height=8.5cm,keepaspectratio]{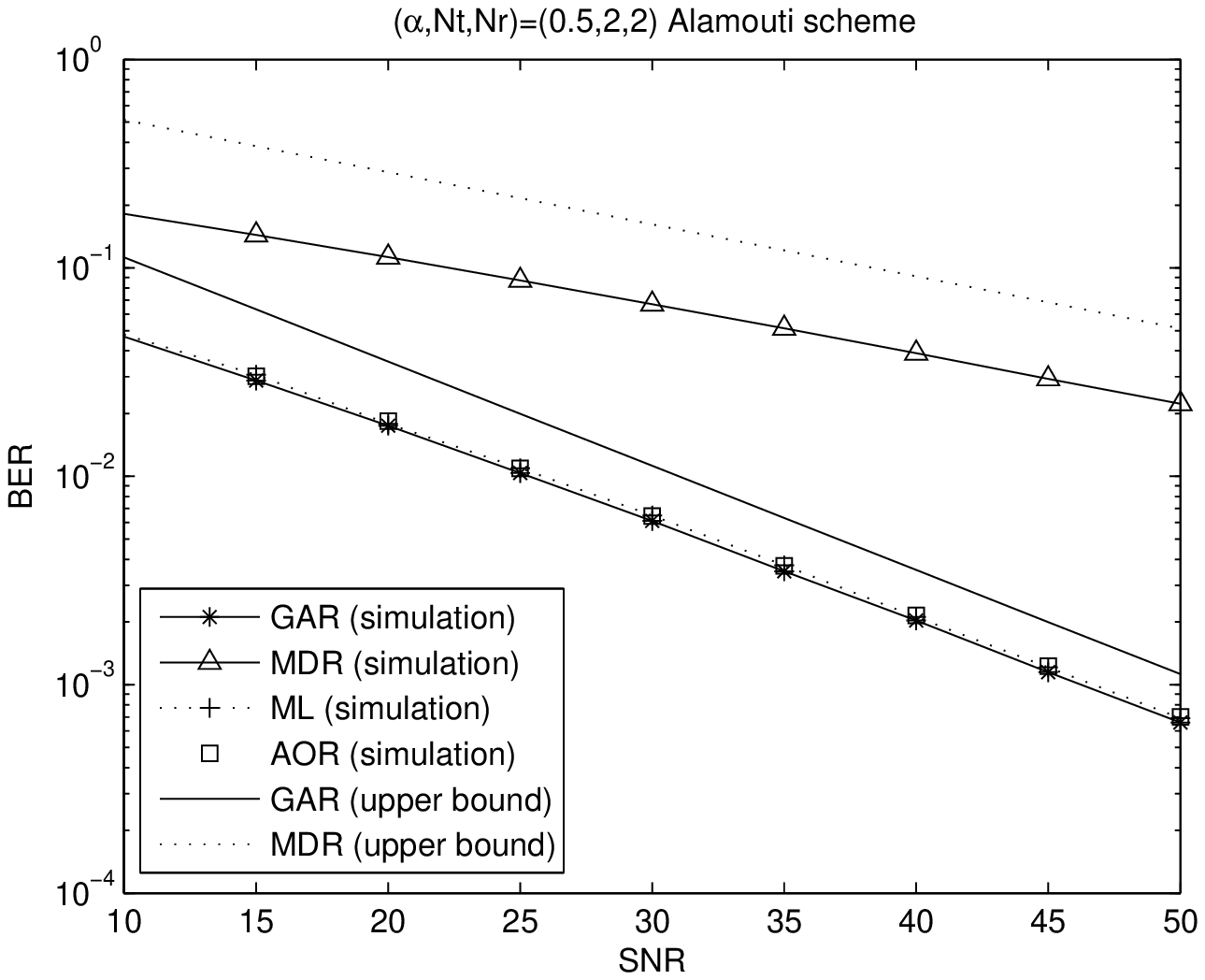}
\caption{Performance comparison of GAR, MDR, ML receiver, and AOR over a channel with highly impulsive noise ($\alpha = 0.5$) with $N_t=2$ and $N_r=2$}\label{fig:Alamouti_2x2_H}
\end{center}
\end{minipage}
\end{figure}

In Fig. \ref{fig:Alamouti_2x2_H}, we show the performance of Alamouti code with $N_t=2,N_r=2$. It is noted that the diversity orders do not change even though the number of receiver antennas increases in accordance with our theoretical result. In this case, the ML receiver and AOR are seen to be within 0.6 dB of the GAR.

\begin{figure}[tb]
\begin{minipage}{1\textwidth}
\centering
\begin{center}
\includegraphics[height=8.5cm,keepaspectratio]{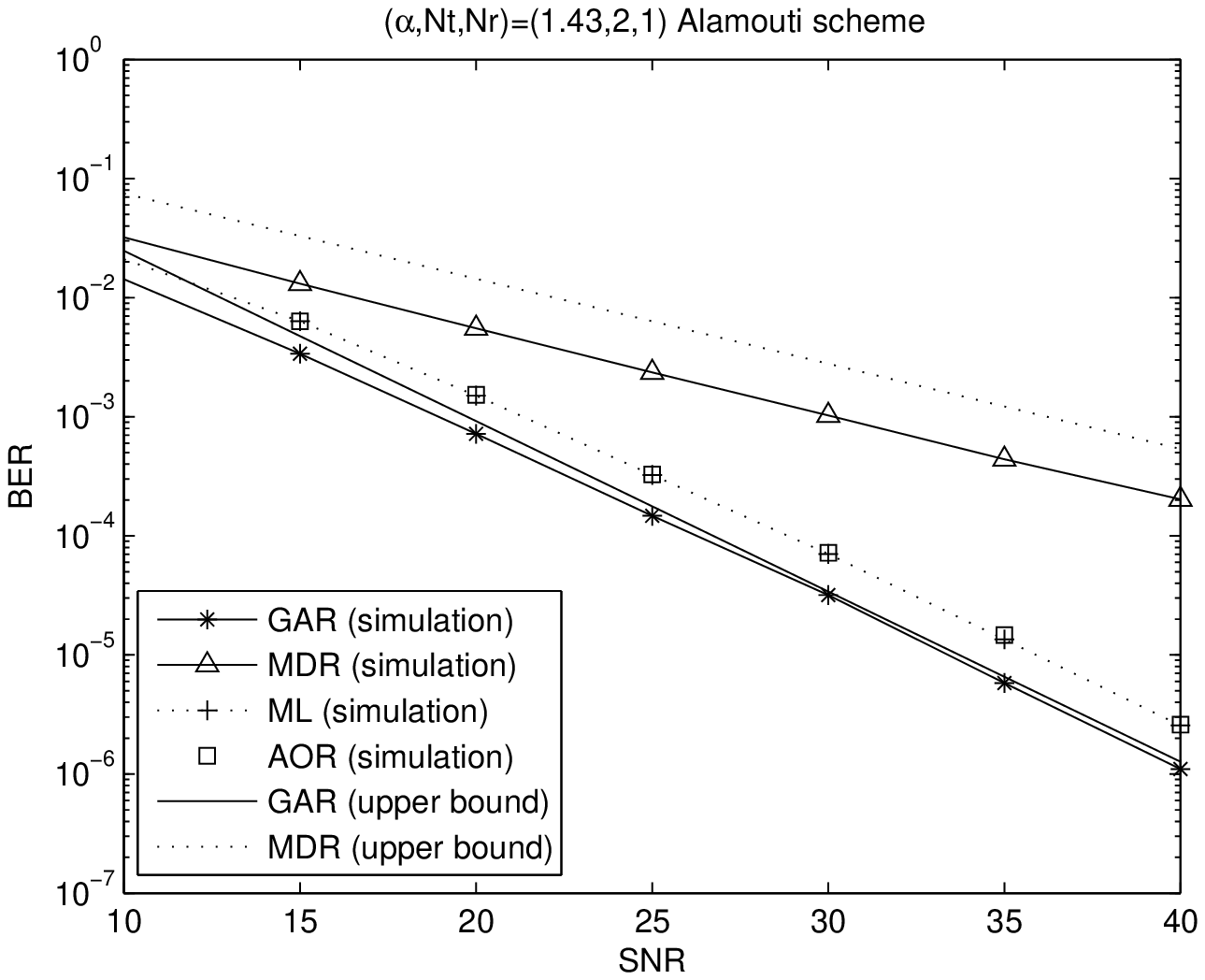}
\caption{Performance comparison of GAR, MDR, ML receiver, and AOR over a channel with moderately impulsive noise ($\alpha = 1.43$) with $N_t=2$ and $N_r=1$}\label{fig:Alamouti_2x1_L}
\end{center}
\end{minipage}
\end{figure}

\begin{figure}[tb]
\begin{minipage}{1\textwidth}
\centering
\begin{center}
\includegraphics[height=8.5cm,keepaspectratio]{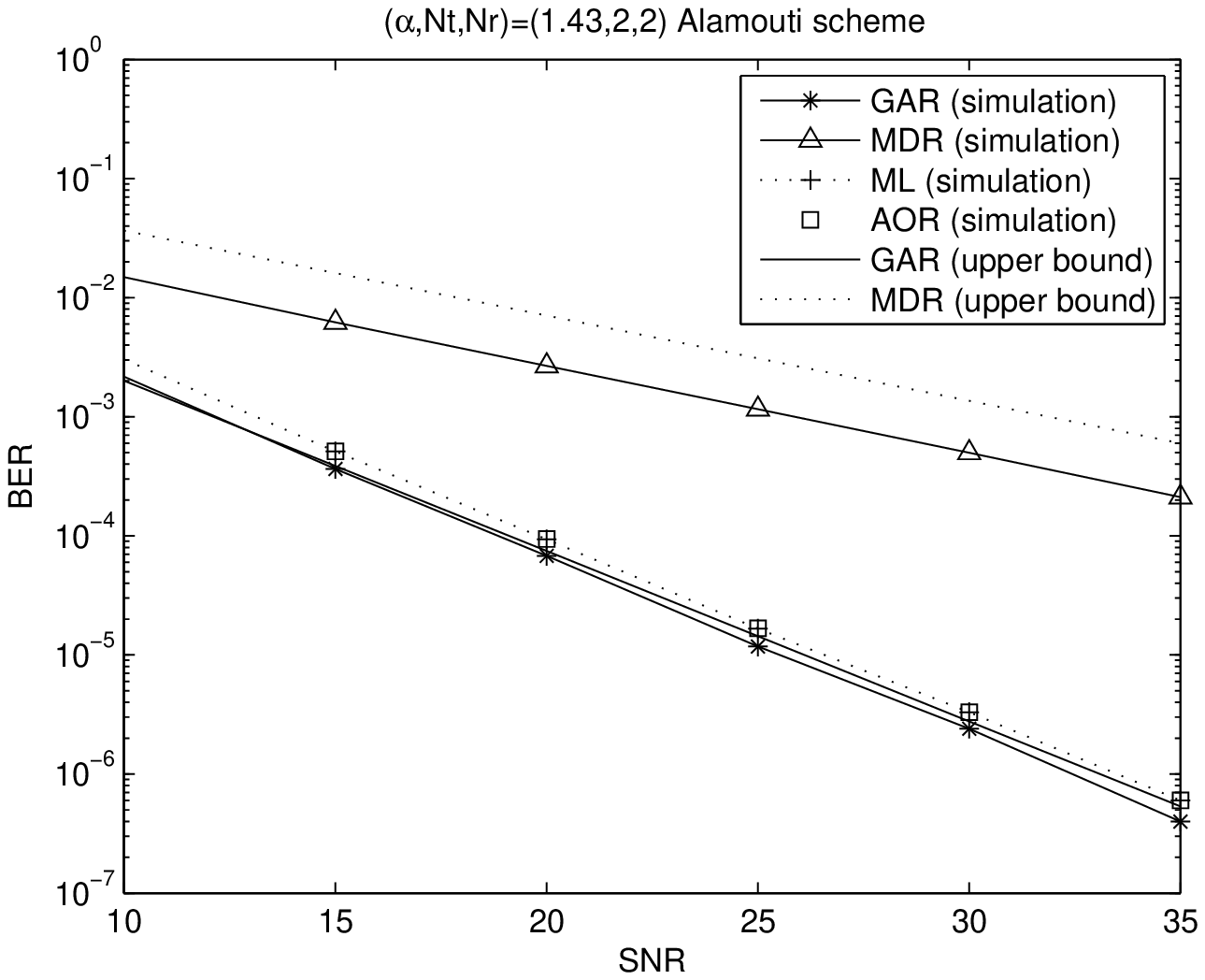}
\caption{Performance comparison of GAR, MDR, ML receiver, and AOR over a channel with moderately impulsive noise ($\alpha = 1.43$) with $N_t=2$ and $N_r=2$}\label{fig:Alamouti_2x2_L}
\end{center}
\end{minipage}
\end{figure}

In the following, we show the performance of Alamouti code over impulsive noise with $\alpha=1.43$. In Fig. \ref{fig:Alamouti_2x1_L}, we show the theoretical and simulated BER with $N_t=2,N_r=1$. The performances with $N_t=2,N_r=2$ are shown in Fig. \ref{fig:Alamouti_2x2_L}. Under the less impulsive noise environment with $\alpha=1.43$, we observe that the diversity orders of GAR and MDR are also $\alpha N_t/2$ and $\alpha/2$ which are in line with our theoretical results. It is also observed the performances for ML receiver and AOR are within 2.5 dB of the GAR at $10^{-3}$ BER, as suggested by Fig. \ref{fig:Alamouti_2x1_L}. In Fig. \ref{fig:Alamouti_2x2_L}, the ML receiver and AOR are seen to be within 1 dB of the GAR.

\subsection{Performance Comparison between Model \Rmnum{1} and \Rmnum{2}}
\begin{figure}[tb]
\begin{minipage}{1\textwidth}
\centering
\begin{center}
\includegraphics[height=8.5cm,keepaspectratio]{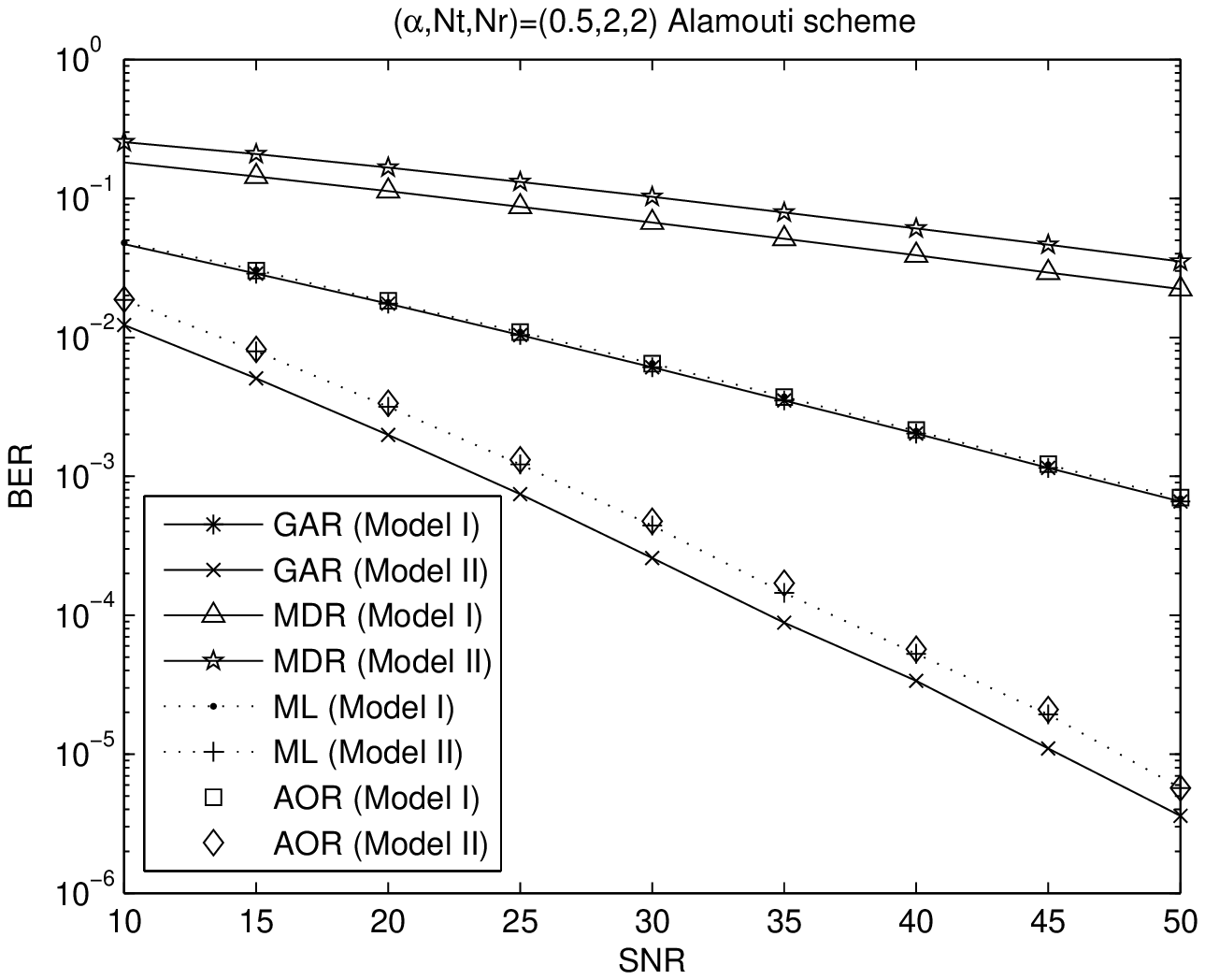}
\caption{Performance comparison of GAR, MDR, ML receiver, and AOR over a channel with highly impulsive noise ($\alpha = 0.5$) with $N_t=2$ and $N_r=2$ under Model \Rmnum{1} and \Rmnum{2}}\label{fig:Alamouti_2x2_H_all}
\end{center}
\end{minipage}
\end{figure}

\begin{figure}[tb]
\begin{minipage}{1\textwidth}
\centering
\begin{center}
\includegraphics[height=8.5cm,keepaspectratio]{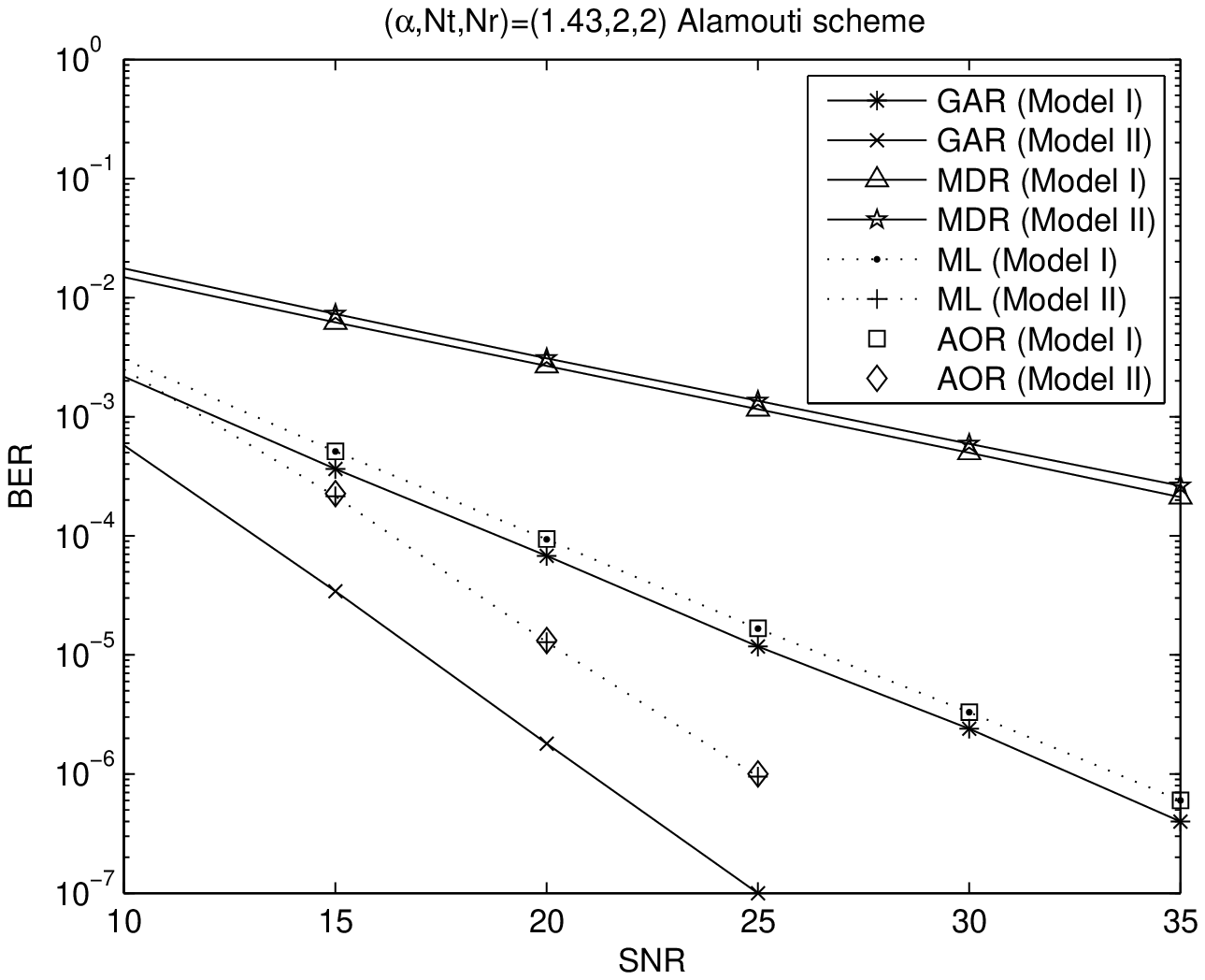}
\caption{Performance comparison of GAR, MDR, ML receiver, and AOR over a channel with moderately impulsive noise ($\alpha = 1.43$) with $N_t=2$ and $N_r=2$ under Model \Rmnum{1} and \Rmnum{2}}\label{fig:Alamouti_2x2_L_all}
\end{center}
\end{minipage}
\end{figure}

In Fig. \ref{fig:Alamouti_2x2_H_all}, we compare the simulated performances of Alamouti code over highly impulsive noise under Model \Rmnum{1} and \Rmnum{2}. Under Model \Rmnum{2}, we can observe that the diversity order of GAR will be larger than that of Model \Rmnum{1}, because additional diversity can be obtained due to the independence of the noise in the space domain. However, the diversity order of MDR does not change even under Model \Rmnum{2}. We can also observe the performance difference for AOR and ML receiver is less than a tenth of a dB. Additionally, we show the simulated performances over impulsive noise with $\alpha=1.43$ under Model \Rmnum{1} and \Rmnum{2} in Fig. \ref{fig:Alamouti_2x2_L_all} where we observe the diversity order of GAR of Model \Rmnum{2} is larger than that of Model \Rmnum{1} and the diversity orders of MDR are always $\alpha/2$. It is also observed the performance difference for AOR and ML receiver is less than a tenth of a dB.

\section{Conclusions}
\label{sec:Conclusions}
In this paper, we considered a S$\alpha$S noise model for MIMO fading channels, and discussed different receivers. In S$\alpha$S noise environments, the diversity order depends on the noise parameter, $\alpha$, and noise correlation model. Under Model \Rmnum{1}, we derived the diversity order for the GAR and MDR. The maximum possible diversity order of GAR is shown to be a benchmark for any receiver, given by $\alpha N_t/2$. The MDR, though simple, is vulnerable to impulsive noise: the diversity order is always $\alpha/2$ regardless the number of antennas. Under Model \Rmnum{2} we have seen that the diversity order for GAR will be larger than that of Model \Rmnum{1}. In contrast, for MDR the diversity order is $\alpha/2$ also for Model \Rmnum{2}.

Since the GAR is impractical to implement, we are motivated to use the ML receiver. However, the ML receiver is computationally complex and requires knowledge of the noise parameters. Thus, we also develop an asymptotically optimal receiver, which performs near optimally at high SNRs and does not require the noise parameters. Since the conventional MDR has poor performance, the usage of the MDR should be avoided in S$\alpha$S noise environments.

\appendices
\section{}
\label{app:GAR}
Since $\lbrace A_i \rbrace_{i=1,...,N_t}$ are i.i.d., each term of the product in \eqref{eqn:GAR_eq_8b} has the same expected value, which results in
\begin{equation}
\label{eqn:A_GAR_eq_2}
E_\mathbf{H,A}P\left(\mathbf{S} \rightarrow \mathbf{S'}\vert\mathbf{A}\right)
=\frac{1}{\pi}\int_0^{\frac{\pi}{2}}\left(E_\mathbf{A}\left[\left(\frac{1}{1+\frac{\rho}{4\sin^2\theta}
\frac{1}{A}}\right)^{N_r}\right]\right)^{N_t}\D{\theta}
\end{equation}
where $A$ represents any of the random variables $A_i$. To simplify the expectation in the RHS of \eqref{eqn:A_GAR_eq_2}, recall that $A \sim S_{\alpha/2}([\cos(\pi\alpha/4)]^{2/\alpha},1,0)$, so that
\begin{equation}
\label{eqn:A_GAR_eq_3}
E_\mathbf{A}\left[\left(\frac{1}{1+\frac{\rho}{4\sin^2\theta}\frac{1}{A}}\right)^{N_r}\right]
=\int_0^{\infty}\left(\frac{4A\sin^2\theta}{4A\sin^2\theta+\rho}\right)^{N_r}f_{\alpha/2}(A)\D{A}.
\end{equation}
The PDF $f_{\alpha/2}(A)$ as suggested by \eqref{eqn:PDF_low_mag} as $A\rightarrow\infty$ is given by
\begin{equation}
\label{eqn:A_GAR_eq_4}
f_{\alpha/2}(A)=\alpha\cos(\pi\alpha/4)C_{\alpha/2}A^{-(1+\alpha/2)}+O\left(A^{-(1+\alpha)}\right) \text{.}
\end{equation}
Substituting \eqref{eqn:A_GAR_eq_4} in \eqref{eqn:A_GAR_eq_3}, we get
\begin{eqnarray}
\nonumber
\lefteqn{\int_0^{\infty}\left(\frac{4A\sin^2\theta}{4A\sin^2\theta+\rho}\right)^{N_r}\left[\alpha C_{\alpha/2} \cos\left(\frac{\pi\alpha}{4}\right) A^{-\left(1+\frac{\alpha}{2}\right)} + O\left(A^{-\left(1+\alpha\right)}\right)\right]\D{A}} \\
\nonumber
&=& \left(\frac{\alpha}{2}\frac{\Gamma\left(\frac{\alpha}{2}\right)\Gamma\left(N_r-\frac{\alpha}{2}\right)}
{\Gamma\left(1-\frac{\alpha}{2}\right)\Gamma(N_r)}\right)\left(\frac{\rho}{4\sin^2\theta}\right)^{-\frac{\alpha}{2}}
+ O\left(\rho^{-\alpha}(4\sin^2\theta)^{\alpha}\right) \\
\label{eqn:A_GAR_eq_5}
&=& \left(\frac{\Gamma\left(1+\frac{\alpha}{2}\right)\Gamma\left(N_r-\frac{\alpha}{2}\right)}
{\Gamma\left(1-\frac{\alpha}{2}\right)\Gamma(N_r)}\right)
\left(\frac{\rho}{4\sin^2\theta}\right)^{-\frac{\alpha}{2}} + O\left(\rho^{-\alpha}\right) \text{.}
\end{eqnarray}
Plugging \eqref{eqn:A_GAR_eq_5} in \eqref{eqn:A_GAR_eq_2}, and using the binomial expansion, we get
\begin{equation}
\label{eqn:A_GAR_eq_6}
P\left(\mathbf{S} \rightarrow \mathbf{S'}\right)=
\frac{1}{\pi}\int_0^{\frac{\pi}{2}}\left[\left(\frac{\Gamma\left(1+\frac{\alpha}{2}\right)\Gamma\left(N_r-\frac{\alpha}{2}\right)}
{\Gamma\left(1-\frac{\alpha}{2}\right)\Gamma(N_r)}\right)\left(\frac{\rho}{4}\right)^{-\frac{\alpha}{2}}\right]^{N_t}\sin^{\alpha N_t}+O\left(\rho^{-\frac{\alpha}{2}(N_t+1)}\right)\sin^{\alpha(N_t-1)}\theta\D{\theta} \text{.}
\end{equation}
Solving the integral in \eqref{eqn:A_GAR_eq_6}, \eqref{eqn:GAR_eq_PEP} follows.

\section{}
\label{app:CG_GAR}
To prove the coding gain, $G_{\rm{GAR}}(N_t,N_r,\alpha)$, is a monotonically decreasing and convex function with respect to $N_t$, we will show a stronger statement which states that the coding gain is a logarithmically completely monotonic (c.m.) function which means that the derivatives of the logarithm satisfy:
\begin{equation}
\label{eqn:Com_mono}
(-1)^n \left(\frac{\partial}{\partial N_t}\right)^{n}\log G_{\rm{GAR}}(N_t,N_r,\alpha) \geq 0
\end{equation}
for $n \in \mathbb{Z}^{+}$. Letting $\alpha N_t/2 = x$ in the coding gain of \eqref{eqn:GAR_eq_PEP}, it suffices to show that
\begin{equation}
\label{eqn:A_CG_GAR_eq_2}
h(x) := \frac{(2\sqrt{\pi})^{\frac{1}{x}}\Gamma(x+1)^{\frac{1}{x}}}{\Gamma\left(x+\frac{1}{2}\right)^{\frac{1}{x}}}
\end{equation}
is a logarithmically c.m. function. Taking logarithm in \eqref{eqn:A_CG_GAR_eq_2}, we get
\begin{equation}
\label{eqn:A_CG_GAR_eq_3}
f(x) = \log h(x) = \frac{1}{x} \log 2\sqrt{\pi}+\frac{1}{x}\log\Gamma(x+1)-\frac{1}{x}\log\Gamma\left(x+\frac{1}{2}\right) \text{.}
\end{equation}
Using Leibnitz' rule, $\left[u(x)v(x)\right]^{(n)} = \sum_{k=0}^{n}\binom{n}{k}u^{(k)}(k)v^{(n-k)}(k)$, in each of the last two terms in \eqref{eqn:A_CG_GAR_eq_3}, we obtain
\begin{eqnarray}
\nonumber
f^{(n)}(x)
&=& (-1)^n\frac{n!}{x^{n+1}}\log2\sqrt{\pi}+
\sum_{k=0}^{n}\binom{n}{k}\left(\frac{1}{k}\right)^{(n-k)}\left[\left[\log\Gamma(x+1)\right]^{(k)}
-\left[\log\Gamma\left(x+\frac{1}{2}\right)\right]^{(k)}\right] \\
\label{eqn:A_CG_GAR_eq_4}
&=& \frac{(-1)^n n!}{x^{n+1}}g(x)
\end{eqnarray}
where
\begin{equation}
\label{eqn:A_CG_GAR_eq_5}
g(x) := \log2\sqrt{\pi}
+\log\Gamma(x+1)-\log\Gamma(x+\frac{1}{2})+\sum_{k=1}^{n}\frac{(-1)^{k}x^{k}}{k!}\psi^{(k-1)}(x+1)-
\sum_{k=1}^{n}\frac{(-1)^{k}x^{k}}{k!}\psi^{(k-1)}\left(x+\frac{1}{2}\right)
\end{equation}
and $\psi^{(n)}(x)$ is the polygamma function as defined follows \cite[pp. 260]{Abramowitz72}:
\begin{equation}
\label{eqn:poly_gamma}
\psi^{(n)}(x) = (-1)^{n+1}\int_{0}^{\infty}\frac{t^n}{1-e^{-t}}e^{-xt}\D{t} \text{.}
\end{equation}
The proof will be complete when we show $g(x) \geq 0$ for $x > 0$ since that would make \eqref{eqn:A_CG_GAR_eq_4} positive. The first derivative of $g(x)$ can be expressed as follows:
\begin{equation}
g'(x) = \frac{(-1)^n x^n}{n!}\left[\psi^{(n)}(x+1)-\psi^{(n)}\left(x+\frac{1}{2}\right)\right] \text{.}
\end{equation}
Using \eqref{eqn:poly_gamma}, we conclude
\begin{eqnarray}
\frac{1}{x^n}g'(x) = \frac{1}{n!}\int_{0}^{\infty}\left(\frac{e^{-\frac{t}{2}}-e^{-t}}{1-e^{-t}}\right)t^n e^{-xt}\D{t} > 0 \text{,}
\end{eqnarray}
since $\left(\frac{e^{-\frac{t}{2}}-e^{-t}}{1-e^{-t}}\right) > 0$ for $t>0$. Thus, the function $g(x)$ is increasing and $g(x)>g(0)>0$ on $(0,\infty)$, which implies $(-1)^n f^{(n)}(x) > 0$ and $n=0,1,2,\dots$. Thus, $h(x)$ is a logarithmically c.m. function. Since a logarithmically c.m. function is also c.m. \cite{Qi04}, $h(x)$ is a c.m. function, which in turn has convex and decreasing functions as a special case. Therefore, it is proved the coding gain is a monotonically decreasing and convex function with respect to $N_t$.

\section{}
\label{app:MDR}
Using \eqref{eqn:CCDF} and order statistics, we can find the PDF of $A_{max}$ as follows:
\begin{equation}
\label{eqn:A_MDR_eq_2}
f_{A_{max}}(x)=\frac{\alpha}{2} T_s 2C_{\alpha/2} \cos\left(\frac{\pi\alpha}{4}\right)x^{-\left(1+\frac{\alpha}{2}\right)}
\left(1-2C_{\alpha/2}\cos\left(\frac{\pi\alpha}{4}\right)x^{-\frac{\alpha}{2}}\right)^{T_s-1}
+O\left(x^{-\frac{\alpha}{2}(T_s+1)-1}\right) \text{.}
\end{equation}
If $T_s \geq 2$ and is equal to $N_t$, plugging \eqref{eqn:A_MDR_eq_2} in \eqref{eqn:MDR_eq_8b} and using binomial expansion, we can get as follows:
\begin{eqnarray}
\nonumber
\lefteqn{\int_0^{\infty}\left(\frac{4x\sin^2\theta}{4x\sin^2\theta+\rho}\right)^{N_rN_t}
\left(\frac{\alpha N_t}{2}\sum_{k=0}^{N_t-1}(-1)^k \left(2C_{\alpha/2} \cos\left(\frac{\pi\alpha}{4}\right)\right)^{k+1}x^{-\left(1+\frac{\alpha}{2}
+\frac{\alpha k}{2}\right)}+O\left(x^{-\frac{\alpha}{2}(N_t+1)-1}\right)\right) \D{x}} \\
\nonumber
&=& \frac{\alpha N_t}{2} \sum_{k=0}^{N_t-1} \left[(-1)^k
\left(\frac{\Gamma\left(\frac{\alpha(1+k)}{2}\right)\Gamma\left(N_rN_t-\frac{\alpha(1+k)}{2}\right)}{\left(\Gamma\left(1-\frac{\alpha}{2}\right)\right)^{(k+1)}\Gamma(N_rN_t)}\right)
\left(\frac{\rho}{4\sin^2\theta}\right)^{-\frac{\alpha(1+k)}{2}}\right]
+O\left(\rho^{-\frac{\alpha(N_t+1)}{2}}\right) \\
\nonumber
&=&\left(\frac{\alpha N_t}{2}
\frac{\Gamma\left(\frac{\alpha}{2}\right)\Gamma\left(N_rN_t-\frac{\alpha}{2}\right)}{\Gamma\left(1-\frac{\alpha}{2}\right)\Gamma(N_rN_t)}\right)
\left(\frac{\rho}{4\sin^2\theta}\right)^{-\frac{\alpha}{2}}
+O\left(\rho^{-\alpha}(4\sin^2\theta)^{\alpha}\right)+\cdots+O\left(\rho^{-\frac{\alpha(N_t+1)}{2}}\right) \\
\label{eqn:A_MDR_eq_4}
&=& \left(\frac{N_t\Gamma\left(1+\frac{\alpha}{2}\right)\Gamma\left(N_rN_t-\frac{\alpha}{2}\right)}
{\Gamma\left(1-\frac{\alpha}{2}\right)\Gamma(N_rN_t)}\right)
\left(\frac{\rho}{4\sin^2\theta}\right)^{-\frac{\alpha}{2}}+O\left(\rho\right)^{-\alpha} \text{.}
\end{eqnarray}
Plugging \eqref{eqn:A_MDR_eq_4} in \eqref{eqn:MDR_eq_8b}, we get
\begin{equation}
\label{eqn:A_MDR_eq_5}
P\left(\mathbf{S} \rightarrow \mathbf{S'}\right)=
\frac{1}{\pi}\int_0^{\frac{\pi}{2}}\left(\frac{N_t\Gamma\left(1+\frac{\alpha}{2}\right)\Gamma\left(N_rN_t-\frac{\alpha}{2}\right)}{\Gamma\left(1-\frac{\alpha}{2}\right)\Gamma(N_rN_t)}\right)
\left(\frac{\rho}{4}\right)^{-\frac{\alpha}{2}}\sin^{\alpha}\theta+O\left(\rho\right)^{-\alpha}\D{\theta} \text{.}
\end{equation}
Solving the integral in \eqref{eqn:A_MDR_eq_5}, \eqref{eqn:MDR_eq_PEP} follows.

\bibliographystyle{IEEEtran}
\nocite{*}
\bibliography{references}
\newpage

\end{document}